\documentclass [a4paper,fleqn, 12pt]{article}
\usepackage{graphicx}
\usepackage[small]{subfigure,epsfig}

\usepackage {amsmath} \usepackage{amssymb} \usepackage{cite}
\newcommand{\cn}

\begin{document}

%\begin{frontmatter}

\title
{Self -- similar solutions of the Burgers hierarchy}

\author
{Nikolay A. Kudryashov}

\date{Department of Applied Mathematics, National Research Nuclear University
MEPHI, 31 Kashirskoe Shosse,
115409 Moscow, Russian Federation}

%\author

%\corauthref{cor}},
%\corauth[cor]{Corresponding author.} \ead{nakudr@gmail.com}

%\address

\maketitle

\begin{abstract}
Self --- similar solutions of the equations for the Burgers hierarchy are presented.

\end{abstract}

%\begin{keyword}

% Point vortices, Special polynomials, Adler -- Moser polynomials

%\PACS 02.30.Hq - Ordinary differential equations

%\end{keyword}

%\end{frontmatter}

\section{Introduction}

The  Burgers hierarchy can be written in the form \cite{Olver, Kudryashov, Kudryashov92,Disine09}
\begin{equation}
\label{BH}u_t+\,\frac{\partial}{\partial x}\,\left(\frac{\partial}{\partial x}+u\right)^n\,u=0, \quad n=0,1,2, \ldots.
\end{equation}

Assuming $n=1$ in Eq. \eqref{BH} we have the Burgers equation
\begin{equation}
\label{B}
u_t+2\,u\,u_{x}+\,u_{xx}=0.
\end{equation}
Eq. \eqref{B} was firstly introduced in \cite{Burgers}. It is well known that this equation can be linearized by means of the Cole-Hopf transformation \cite{Hopf51,Cole50, Kudryashov09am}. Exact solutions of Eq.\eqref{B} were considered in many papers (see, for example, \cite{Rosenblatt68, Benton66, Malfliet, Fahmy}).

Assuming $n=2$ in Eq. \eqref{BH} we obtain the Sharma - Tasso - Olver equation
\begin{equation}
\label{S}
u_t+\,u_{xxx}+3\,u_{x}^{2}+3\,\,u\,u_{xx}+3\,\,u^{2}\,u_{x}=0.
\end{equation}
The Sharma - Tasso - Olver equation was derived in \cite{Olver, Tasso}. Some exact solutions of this equation were presented in \cite{Hereman, Yang, Shang, Lou, Kudryashov08, Kudryashov09a, Kudryashov09b, Kudryashov09c}.

At $n=3$ and $n=4$ we obtain the following fourth and fifth order partial differential equations
\begin{equation}\begin{gathered}
\label{FM}
u_{t} +\,u_{xxxx}  +10\,
u_{x} u_{xx}  +4\,u
u_{xxx} +12\,u   u_{x}^{2}+\\+6\,
  u  ^{2} u_{xx}  +4\,u^{3}\,u_{x}=0,
\end{gathered}\end{equation}
\begin{equation}\begin{gathered}
\label{FFM}
u_{t} +\,u_{xxxxx}  +10\,
u_{xx}^{2}+15\, u_{x}
u_{xxx}  +5\,u u_{xxxx}  +15\,
 u_{x}^
{3}+
\\+50\,u u_{x} u_{xx}  +10\, u^{2}u_{xxx}
  +30\, u^{2} u_{x}^{2}+10\,u^{3} u_{xx}  +5\, u ^{4}u_{x} =0.
\end{gathered}\end{equation}

Assuming
\begin{equation}
\label{Tran} x=L\,x^{'}, \quad u=C_0\,u^{'}, \quad t=T\,t^{'},
\end{equation}
we have that Eq.\eqref{BH} is invariant under the dilation group in the case
\begin{equation}
\label{Tran1} C_0\,L=1, \quad T=L^{n+1}.
\end{equation}
Assuming $C_0=e^{-a}$  in \eqref{Tran1}, we obtain the delation group for the Burgers hierarchy \eqref{BH} in the form
\begin{equation}
\label{Tran2} u^{'}=e^{-a}\,u,\quad  x^{'}=e^{a}\,x,\,\quad t^{'}=e^{a(n+1)}\,t.
\end{equation}

From transformations \eqref{Tran2} we have two invariants for Eq.\eqref{BH}
\begin{equation}
\label{Tran3} I_1=u\,t^{\frac{1}{n+1}}=u^{'}\,(t^{'})^{\frac{1}{n+1}},\qquad I_2=\frac{x}{t^{\frac{1}{n+1}}}=\frac{x^{'}}{(t^{'})^{\frac{1}{n+1}}}.
\end{equation}

Therefore we look for the solutions of the Burgers hierarchy taking into account the variables
\begin{equation}
\label{BH2}u(x,t)=\frac{A}{t^{\frac{1}{n+1}}}\,f(z), \quad z=\frac{B\,x}{t^{\frac{1}{n+1}}}.
\end{equation}

Substituting \eqref{BH2} into \eqref{BH} we obtain the equation for $f(z)$ at
\begin{equation}
\label{BH3} A=B=\frac{1}{(n+1)^{\frac{1}{n+1}}}.
\end{equation}
in the form
\begin{equation}
\label{BH5}\left(\frac{d}{dz}+f\right)^n f-z\,f+\beta=0,
\end{equation}
where $\beta$ is the constant of integration.

Solving Eq.\eqref{BH5} we obtain solutions of the Burgers hierarchy in the form
\begin{equation}
\label{BH5a}u(x,t)=\frac{1}{\left(n\,t+t\right)^{\frac{1}{n+1}}}f(z), \quad z=\frac{x}{\left(n\,t+t\right)^{\frac{1}{n+1}}}.
\end{equation}
Let us study the solutions of nonlinear ordinary differential equation \eqref{BH5}.

\section{Exact solutions of equation\eqref{BH5}}

First of all let us prove the following lemma.

\textbf{\emph{Lemma 1.}} \emph{Equation \eqref{BH5} can be transformed to the linear equation of $(n+1)$ - th order by means of transformation}
\begin{equation}
\label{BH6}f=\frac{\psi_{z}}{\psi}.
\end{equation}

\textbf{\emph{Proof.}} The proof of this lemma can be given by means of the mathematical induction method.

Using the transformation \eqref{BH6} we have
\begin{equation}
\label{BH7}\left(\frac{d}{dz}+f\right)\,f=\frac{\psi_{zz}}{\psi},\quad \left(\frac{d}{dz}+f\right)^2\,f=\frac{\psi_{zzz}}{\psi}
\end{equation}

Assuming that there is equality
\begin{equation}
\label{BH7a} \left(\frac{d}{dz}+f\right)^k\,f=\frac{\psi_{k+1,z}}{\psi},\quad
\psi_{k+1,z}=\frac{d^{k+1}\psi}{d\,z^{k+1}}.
\end{equation}

Differentiating Eq.\eqref{BH7a} with respect to in $z$ we have
\begin{equation}
\label{BH7b} \frac{d}{dz}\,\left(\frac{d}{dz}+f\right)^k\,f=\frac{\psi_{k+2,z}}
{\psi}-\frac{\psi_z\,\psi_{k+1,z}}{\psi^2}.
\end{equation}

From Eq.\eqref{BH7b} we obtain the equality
\begin{equation}
\label{BH7c} \left(\frac{d}{dz}+f\right)^{k+1}f=\frac{\psi_{k+2,z}}
{\psi}.
\end{equation}

Therefore we obtain the formula
\begin{equation}
\label{BH7d} \left(\frac{d}{dz}+f\right)^n f=\frac{\psi_{n+1,z}}{\psi}.
\end{equation}

Taking this formula into account we have the equality
\begin{equation}
\label{BH7d} \left(\frac{d}{dz}+f\right)^n f-z\,f+\beta=\frac{1}{\psi}\,(\psi_{n+1,z}-z\,\psi_z+\beta\,\psi).
\end{equation}

As result of this lemma we obtain that solutions of Eq. \eqref{BH5} can be found by the formula  \eqref{BH6}, where $\psi(z)$ is the solution of the linear equation
\begin{equation}
\label{BH8}\psi_{n+1,z}-z\,\psi_z+\beta\,\psi=0,
\end{equation}

Let us consider the partial  cases. Assuming $\beta=0$ in Eq.\eqref{BH8} we have
\begin{equation}
\label{BH88}\psi_{n+1,z}-z\,\psi_z=0.
\end{equation}
Denoting $\psi_z=y$ we obtain
\begin{equation}
\label{BH89}y_{n,z}-z\,y=0.
\end{equation}

In the case $n=1$ we get solution of Eq.\eqref{BH89} in the form
\begin{equation}
\label{BH8a}y(z)=C_2\,e^{-\frac{z^2}{2}}.
\end{equation}
The general solution of Eq.\eqref{BH89} can be written as
\begin{equation}
\label{BH8b}\psi(z)=C_3+C_2\,\int_{0}^{z}\,e^{-\frac{\xi^2}{2}}d\xi,
\end{equation}
where $C_2$ and $C_3$ are arbitrary constants.
In the case $n=2$  we obtain
the general solution of Eq.\eqref{BH89} in the form
\begin{equation}
\label{BH8d}y{(z)}=C_4\,\sqrt{z}J_{\frac13}\left(\frac23\,z^
{\frac32}\right)+C_5\,\sqrt{z}\,Y_{\frac13}\left(\frac23\,z^{\frac32}\right),
\end{equation}
where $J_{\frac13}$ and $Y_{\frac13}$ are the Bessel functions.

In the case $n>2$ solution of Eq.\eqref{BH89} has $n$ solutions
\begin{equation}
\label{BH8m}y_j{(z)}=z^{j-1}\,E_{n,1+\frac1n, 1+\frac{j}{n}}(z^{n+1}),\quad j=1,2,\ldots,n,
\end{equation}
where $E_{n,m,l}$ is a Mittag - Leffler type special function defined by \cite{Polyanin03};
\begin{equation}
\label{BH8n}E_{n,m,l}(z)=1+\sum_{k=1}^{\infty}b_k\,z^k,\quad b_k=\prod_{s=0}^{k-1}\frac{\Gamma\left(n(ms+l)+1\right)}{\Gamma{\left(n(ms+l+1)+1\right)}}
\end{equation}

In the case $\beta\neq0$ solutions of Eq.\eqref{BH89} can be referred to the type of the Laplace equations\cite{Polyanin07}. There are partial solutions $\psi(z)=-z^m$ of Eq.\eqref{BH8} at $\beta=m$, where $0<m\leq n$ is integer. In the general case solutions of equations \eqref{BH8} can be found using the Laplace transformation or taking the expansions in the power series into account.

For a example let us solve the Cauchy problem for linear ordinary differential equation \eqref{BH8} at $\beta=-1$. We have the following problem
\begin{equation}\begin{gathered}
\label{BH9}\psi_{n+1,z}-z\,\psi_z-\,\psi=0,\\
\\
\psi(z=0)=b_0,\quad \psi_z(z=0)=b_1, \ldots, \psi_{n-2,z}=b_{n-2}\quad \psi_{n-1,z}=b_{n-1}.
\end{gathered}\end{equation}

Substituting
\begin{equation}
\label{BH10} \psi(z)=\sum_{m=0}^{\infty}a_m\,z^{m}
\end{equation}
into Eq.\eqref{BH9}, we obtain the solution in the form
\begin{equation}\begin{gathered}
\label{BH11}\psi(z)=a_0\,\sum_{k=0}^{\infty}\frac{z^{nk}\,\prod_{j=0}^{k}{(n\,j+1})}{(nk+1)!}
+a_1\,\sum_{k=0}^{\infty}\frac{z^{nk+1}\,\prod_{j=0}^{k}{(n\,j+2})}{(nk+2)!}+\\
\\
+2\,a_2\,\sum_{k=0}^{\infty}\frac{z^{nk+2}\,\prod_{j=0}^{k}{(n\,j+3})}{(nk+3)!}+\ldots+\\
\\
+(n-2)!\,a_{n-2}\,\sum_{k=0}^{\infty}\frac{z^{nk+n-2}\,\prod_{j=0}^{k}{(n\,j+n-1})}{(nk+n-1)!}+\\
\\+
(n-1)!\,a_{n-1}\,\sum_{k=0}^{\infty}\frac{z^{nk+n-1}\,\prod_{j=0}^{k}{(n\,j+n})}{(nk+n)!}.
\end{gathered}\end{equation}

The value of coefficients $a_0$, $a_1$, $a_2, \ldots, a_{n-2}$  and $a_{n-1}$ are determined by the initial values $b_0$, $b_1$, $b_2, \ldots, b_{n-2}$ and $b_{n-1}$. We have
\begin{equation}
\label{BH11a} a_0=b_0,\quad a_1=b_1,\quad a_2=\frac{b_2}{(2!)^2}, \ldots, a_{n-1}=\frac{b_{n-1}}{((n-1)!)^2}.
\end{equation}

Let us present the partial cases of solution for equation \eqref{BH9}. In the case $n=3$ we have solution in the form
\begin{equation}\begin{gathered}
\label{BH12}\psi(z)=a_0\,\sum_{k=0}^{\infty}\frac{z^{3k}\,\prod_{j=0}^{k}{(3\,j+1})}{(3k+1)!}
+a_1\,\sum_{k=0}^{\infty}\frac{z^{3k+1}\,\prod_{j=0}^{k}{(3\,j+2})}{(3k+2)!}+\\
\\
+2\,a_2\,\sum_{k=0}^{\infty}\frac{z^{3k+2}\,\prod_{j=0}^{k}{(3\,j+3})}{(3k+3)!}.
\end{gathered}\end{equation}

Assuming $n=4$ we obtain
\begin{equation}\begin{gathered}
\label{BH11}\psi(z)=a_0\,\sum_{k=0}^{\infty}\frac{z^{4k}\,\prod_{j=0}^{k}{(4\,j+1})}{(4k+1)!}
+a_1\,\sum_{k=0}^{\infty}\frac{z^{4k+1}\,\prod_{j=0}^{k}{(4\,j+2})}{(4k+2)!}+\\
\\
+2\,a_2\,\sum_{k=0}^{\infty}\frac{z^{4k+2}\,\prod_{j=0}^{k}{(4\,j+3})}{(4k+3)!}
+6\,a_{3}\,\sum_{k=0}^{\infty}\frac{z^{4k+3}\,\prod_{j=0}^{k}{(4\,j+4})}{(4k+4)!}.
\end{gathered}\end{equation}

One can show that these power  series are conversed for any values $z$. Therefore self-similar solutions of equations for the Burgers hierarchy are found after substitution \eqref{BH11} into formula \eqref{BH6}.

Author is grateful to Andrey Polyanin for useful discussion of nonlinear differential equation Eq.\eqref{BH8}.


\begin{thebibliography}{99}


\bibitem{Olver} Olver P.J., Evolution equations possessing infinitely many symmetries, J. Math. Phys. 18 (1977), 1212 -1216

\bibitem{Kudryashov} \textit{ Kudryashov N.A.}, Analitical theory of
nonlinear differential equations, Moskow - Igevsk, Institute of
computer investigations, 2004, (in Russian)

 \bibitem{Kudryashov92} \textit{Kudryashov N.A.},  Partial differential equations with
solutions having movable forst - order singularities, Physics
Letters A, 1992, 169:  237 - 42

\bibitem{Disine09} Kudryashov N.A., Sinelshchikov D.I., Exact solutions for equations of the Burgers hierarchy, Appl. Math. Comput., (2009), doi:10.1016/j.amc.2009.06.010

\bibitem{Burgers} J.M. Burgers, A mathematical model illustrating the theory of turbulance, Advances in Applied Mechanics.1 (1948) 171-199.

\bibitem{Hopf51} E. Hopf, The partial differential equation $u_t+u\,u_x=u_{xx}$,
Communs. Pure Appl. Math. 3 (1950) 201-230.

\bibitem{Cole50} J.D. Cole, On a quasi-linear parabolic equation occuring in aerodynamics
 Quart. Appl. Math. 9 (1950) 225-236.

 \bibitem{Kudryashov09am} N.A. Kudryashov, M.V. Demina, Traveling wave solutions of the generalized nonlinear evolution equations, Applied Mathematics and Computation, 210,
(2009), 551--557

\bibitem{Rosenblatt68}  M. Rosenblatt, Remark on the Burgers equation,
 Phys. Fluids. 9 (1966) 1247-1248.

\bibitem{Benton66}  E.R. Benton,  Some New Exact, Viscous, Nonsteady Solutions of Burgers' Equation,
 J. Math. Phys.  9 (1968) 1129-1136.

\bibitem{Malfliet} W. Malfliet, Approximate solution of the damped Burgers equation, J. Phys. A. 26 (1993) L723-L728.

\bibitem{Fahmy} E. S. Fahmy, K. R. Raslan, H. A. Abdusalam, On the exact and numerical solution of the time-delayed Burgers equation, International Journal of Computer Mathematics. 85 (2008) 1637-1648

\bibitem{Tasso} A. S. Sharma , H. Tasso,  Connection between wave envelope and explicit solution of a nonlinear
dispersive equation. Report IPP 6/158. 1977.


\bibitem{Hereman} W. Hereman, P.P. Banerjee, A. Korpel, G. Assanto,
A. Van Immerzeele, A. Meerpoel, Exact solitary wave solutions of non-linear evolution and wave
equations using a direct algebraic method, J. Phys. A Math. Gen. 19 (1986) 607-628.

\bibitem{Yang} Z. J. Yang, Travelling wave solutions to nonlinear evolution and wave
equations, J. Phys. A Math. Gen. 27 (1994) 2837-2855.

\bibitem{Shang} Y. Shanga, J. Qina, Y. Huangb, W. Yuana, Abundant exact and explicit solitary wave and periodic wave solutions to the Sharma–--Tasso–--Olver equation, Applied Mathematics and Computation. 202 (2008) 532-538.

\bibitem{Lou} S. Wang, X. Tang, S.-Y. Lou, Soliton fission and fusion: Burgers equation and
Sharma--–Tasso--–Olver equation. Chaos, Solitons and Fractals. 21 (2004) 231-239.

\bibitem{Kudryashov08} N.A. Kudryashov, N.B. Loguinova  Extended simplest equation method for nonlinear differential equations , Applied Mathematics and Computation. 205 (2008) 396 - 402.

\bibitem{Kudryashov09a} N.A. Kudryashov, N.B. Loguinova, Be carefull with Exp - function
method, Commun Nonlinear Sci Numer Simulat, 14 (2009), 1881 - 1890

\bibitem{Kudryashov09b} N.A. Kudryashov, On "new travelling wave solutions" of the KdV and
the KdV - Burgers equations, Commun Nonlinear Sci Numer Simulat, 14
(2009), 1891--- 1900

\bibitem{Kudryashov09c} N.A. Kudryashov,  Seven common errors in finding exact solutions
of nonlinear differential equations, Commun Nonlinear Sci Numer Simulat, 14 (2009), 3507 - 3509

\bibitem{Polyanin03} A.D. Polyanin, V.F. Zaittsev,  Handbook of Exact Solutions for Ordinary Differential Equations, Chapman and Hall/CRC Press, 2003, 689 - 733

\bibitem{Polyanin07} A.D. Polyanin and A.V. Manzhirov, Handbook of Mathematics  for Engineers and Scientists, Chapman and Hall/CRC Press, 2007, 518 - 522








\end{thebibliography}
\end{document}